\documentstyle[12pt,aps,epsfig]{revtex}
\begin{document} 
\draft
\title{Proton stopping in C+C, d+C, C+Ta and d+Ta  collisions at 4.2A GeV/c}
\author{ Lj. Simi\'c \thanks{E-mail: simic@atom.phy.bg.ac.yu}
 and M. Kornicer  \thanks{E-mail: kornicer@atom.phy.bg.ac.yu} } 
\address{Institute of Physics, P.O. Box 57, 11001 Belgrade, Yugoslavia}
\maketitle
\begin{abstract} 

       The shape of proton rapidity distributions is analysed in terms 
 of their Gaussian 
 components, and the average rapidity loss is determined in order 
 to estimate the amount of  
 stopping in C+C, d+C, C+Ta and d+Ta collisions at 4.2A GeV/c. Three Gaussians 
 correspond to the nuclear transparency and describe well 
 all peripheral and also C+C 
 central collisions. Two-component shape is obtained in  case of 
 d+C and C+Ta  central 
 collisions. Finally one Gaussian, found in  d+Ta central 
 collisions, corresponds to the  full stopping. 
 The calculated values of  the average rapidity loss 
 support  the qualitative 
 relationship between the number of Gaussian  components and 
 the  corresponding 
 stopping power. It is observed in central  collisions that  
 the average rapidity loss 
 increases with the ratio of the number of target and the number of 
 projectile participants.

\end{abstract}

\clearpage

      Baryon rapidity distribution can be used to  estimate the nuclear 
 stopping achieved in a 
 collision  i.e. the fraction of  the projectile kinetic energy deposited 
 in the reaction volume  
 and converted into  other degrees of freedom. When full stopping  
 is achieved, the rapidity 
 distribution  has a maximum at midrapidity with the width determined  
 by thermal and 
 possibly  hydrodynamic motion. If the nuclei are highly transparent to each  
 other, the final  
 proton distribution exhibits peaks at or near target and beam  rapidities, 
 and a sparse  
 population of the central rapidity region. Study of the proton  
 rapidity distribution as  a 
 function of the system size and of the impact parameter has  shown 
 that peripheral  
 collisions resemble the p+p data \cite {Vid1,Vid2}, where a small 
 stopping is achieved.  
 Central collisions of light nuclei  (Si+Al, S+S) reveal additional  
 stopping with a fairly  
 wide and flat rapidity distribution \cite {Vid1,Vid2,Abbo,Bar,Bach}.  
 Larger, but  not 
 complete,  stopping is observed  in central collisions of heavier systems, 
 such as Au+Au at 
 AGS  \cite {Vid1,Lac}  and Pb+Pb at  SPS energies \cite {Jon,Wien,Jac}. 
 Proton rapidity 
 distribution as a function of the beam energy shows  that, at Bevelac energy 
 the complete stopping 
 results in a single fireball, while  at higher energies the distribution 
 broadens and changes 
 its shape \cite {Wien,Gut}.  It is found that for symmetric systems 
 the  stopping is larger 
 for heavier systems and lower energies \cite {Jac}. In order to 
 estimate the amount of 
 stopping,  the corresponding  average rapidity loss is used. 
 A review of the average  
 rapidity loss for p+p and A+A collisions at beam momenta 1-2, 11.6, 14.6 
 and 200 GeV/c per 
 nucleon can be found in \cite{Vid2,Hon}. 
 
      Here, the shape of proton rapidity distributions in peripheral  
 and central  C+C, d+C, 
 C+Ta and d+Ta collisions at 4.2A GeV/c, is analysed in 
 terms of  corresponding  
 Gaussian components. The corresponding  average  rapidity loss is 
 calculated and its  
 dependence on the  collision centrality and  system size is discussed.
 The study is based 
 on a sample of 7327 C+C, 6735 d+C, 1989  C+Ta and 1475   d+Ta events. 
 The data are 
 obtained with the 2-m propane bubble  chamber, exposed at the JINR,  Dubna 
 synchrophasotron with $^{12}$C or $^{2}$d ion beam of 4.2A GeV/c.  
 Additionally the 
 $^{181}$Ta  target, consisting of the three foils 
 (1 mm thick and 93 mm apart),  was  
 placed  inside the chamber working in the 1.5 T magnetic field.  
 This allows the study of  
 inelastic interactions with carbon as well as with tantalum target.  
 The characteristics  of 
 the chamber allow precise determination of multiplicity  
 and momentum of all  charged 
 particles, as well as identification of all negative particles  
 and positive particles  with 
 momenta less than 0.5 GeV/c. The latter are classified either  
 as protons or $\pi^+$  
 mesons according to their ionisation density and range.  
 All positive particles above  0.5 
 GeV/c are taken to be protons, and due to this  the admixture of $\pi^+$, 
 of about  (10- 15)\%, is subtracted, using the 
 number of $\pi^-$ mesons with $p>0.5$ GeV/c 
 as follows:
 $n_{p}=n_{+}-n_{\pi^+}(p\leq0.5 GeV/c)-kn_{\pi^-}(p>0.5 GeV/c)$, where 
 $n_+$ denotes the number of single positively charged particles, $k=1$ for
 collisions of isoscalar nuclei, and $k=0.82$ for collisions 
 with tantalum nuclei. This last value is less  than one since it 
 takes into account 
 the proton deficit in tantalum nuclei and consequently also $\pi^+$ deficit.
 From the ratio $w=n_{p}/n_{+}$ for each momentum interval 
 we determine the weight of protons which we further use when calculating
 distributions of other kinematical variables.  
 From the resulting number of  protons, the 
 projectile spectator protons (with momenta $p>3$ GeV/c  
 and emission  angle $\theta <4^0$), and target spectator 
 protons  (with momenta $p<0.3$ GeV/c) are further  
 subtracted. The resulting number of participant protons still  
 contains some 17\% of  
 deuterons (with momenta $p>0.48$ GeV/c), and 11\% of tritons 
 with  (momenta $p>0.65$ 
 GeV/c). The centrality selection   is performed according to the 
 number  of participant 
 protons $n_p$. The   
 events with $n_{p} < \langle n_{p}\rangle$  i.e. with $n_p \le2$ in d+C,
 $n_p \le3$ in d+Ta,  $n_p \le4$ in  
 C+C and  with $n_p \le 10$ in C+Ta 
 collisions are 
 classified as peripheral. The selected peripheral events correspond to
 50-60\% cross section cut. In order to show that the 
 criterion of peripherality is not affecting the results, we divide 
 the peripheral events into two subgroups (which statistics allows)
 corresponding to different 
 cuts in $n_p$.
 The   events with the largest 
 multiplicity  of participant  protons, 
 corresponding to (3-5)\% cross section cut, are classified as central. 

      Fig. 1 shows proton rapidity distributions for peripheral and  
 central C+C,  d+C, C+Ta 
 and d+Ta collisions. For all peripheral collisions,  two distinct peaks  
 consisting of target 
 and projectile protons can be seen, indicating  the small 
 amount of  stopping. For 
 symmetric collision system, the peaks are  symmetrically positioned  
 around center of mass 
 rapidity  $(y_{cm}=1.1)$ and distribution  resembles the pp data. 
 For  asymmetric 
 collision systems the target peak is more prominent  then projectile peak.  
 Additional stopping is achieved with increasing centrality. 
 In case of symmetric collisions this leads 
 to appearance  of a flat plateau in distribution.  
 In case of asymmetric  collisions 
 the projectile-like peak disappears so that only  the target-like 
 peak with  long tail towards 
 the projectile rapidity remains.  In C+C, d+C, and C+Ta  collisions, 
 the shape of $dN/dy$ 
 distribution remains largely unchanged  for a centrality  
 cut below 10\%. On the other 
 hand, in d+Ta collisions the  long tail of  the rapidity distribution 
 gradually diminishes 
 with decreasing  centrality cut, and  completely disappears 
 below 3\% so that only the 
 target-like peak  remains.  

      Fig. 1 also shows that the shape of rapidity distributions  
 is best reproduced by  a 
 sum of several ($ \le 3$) Gaussians. The exact values of the  
 best fit parameters  (peak 
 positions and  widths) can be easily 
 acquired from the 
 corresponding  figures.  According to the peak position, 
 these Gaussians are related  to the 
 central region  (the peak close to the participant center-of-mass rapidity),
 and the target 
 and  projectile fragmentation regions. Three Gaussians are  
 obtained in all  peripheral 
 collisions, and C+C central collisions. With the finer cut in peripheral 
 events, the rapidity distributions 
 retain the three Gaussian structure so that this 
 structure is not the result of the impact parameter superposition.
 Two Gaussians,  
 central and target-like, are found 
 in d+C and C+Ta central collisions.  Finally, one target-like Gaussian,  
 is obtained in 
 d+Ta central collisions. The number of Gaussians  depends on the  
 collision centrality 
 and/or stopping. Three Gaussians correspond to nuclear  
 transparency while single 
 Gaussian describes full stopping scenario.  Since it is  possible 
 to associate each Gaussian 
 component in rapidity distribution  with single  isotropic thermal 
 source (fireball) \cite {Sch,Braun}, one arrives at simple  physical 
 picture consisting  of several fireballs in 
 relative motion. In case of symmetric  collisions (such as C+C)  
 there will be, for a given 
 impact parameter, an overlap between  the target and the  projectile. 
 After collision, the 
 overlapping regions fuse together  and partially come  to rest 
 in the center of mass frame. 
 The halted region forms  subsequently the central  fireball. 
 The two broken off parts 
 continue in their paths after  collision with  reduced momentum 
 and with smaller 
 amount of the initial collision  energy. These eventually create 
 the two fragmentation  fireballs. 
 In central collisions of an asymmetric  system (such as  d+C, C+Ta), 
 smaller  projectile 
 nucleus  penetrates right trough the target forming in 
 the process the early  stage of the 
 central  fireball. 
 The target and/or projectile participant 
 nucleons left  behind 
 lead  subsequently to the only one, slower, fragmentation fireball. 
 Thus one expects, for  
 asymmetric  system and central collisions only two fireballs.  
 When  $A_p \ll A_t$, 
 as in case  of d+Ta, the lighter projectile is not able to penetrate 
 the heavy  target (which 
 looks  black for projectile) and consequently only one, target-like,  
 fireball appears. 

      In order to quantify the amount of stopping, the average  
 rapidity loss (or  mean 
 rapidity shift) of projectile protons,  
 $\langle \delta y \rangle= y{_p}- \langle y \rangle$, is 
 introduced.  Here, $y_p =2.2$  is the beam rapidity, 
 and $\langle y \rangle$ is the average 
 rapidity of  the projectile protons. In all  peripheral 
 and C+C central collisions all protons 
 with rapidity  $y > y_m = y{_p}/2$ are taken  as projectile protons, 
 and averaging is 
 performed from $y_m$  to $y_p$. In central d+C,  
 C+Ta and d+Ta collisions, the 
 averaging is taken from $y_p$ to a  rapidity $y_{low}$  
 determined so that the integral of 
 the proton rapidity distribution  is equal to the  number of 
 interacting projectile protons 
 \cite {Vid2}. When the target  and projectile protons  mix strongly, 
 the average rapidity 
 loss determined in this manner  is a lower limit.  
 The resulting $\langle \delta y \rangle$ 
 values for peripheral and  central collisions are summarised in  Table 1. 
 For peripheral  collisions, average rapidity loss is approximately  
 independent of  the collision system. In 
 central collisions,  $\langle \delta y \rangle$  has the smallest  
 value in C+C (somewhat 
 larger than in peripheral C+C) and the largest value in d+Ta  collisions. 
 These values for the 
 average rapidity loss support the   qualitative  relationship between 
 the number of 
 Gaussian components in the  rapidity spectra  and the corresponding 
 stopping power. The 
 quantification of the  stopping power via   
 $\langle \delta y \rangle$  yields additional 
 information.  Evidently, the carbon nucleus as a target is less effective  
 than the tantalum nucleus in slowing 
 down the incident projectile nucleons.  For a fixed  target, 
 the stopping depends on the size 
 of the projectile.  Generally, we find that   
 $\langle \delta y \rangle$  increases with 
 $A_{t}/A_{p}$  or,  more correctly, with   $[A_{t}-
 (A_{t}^{2/3}-A_{p}^{2/3})^{3/2}]/A_{p}$.  Here, the numerator represents 
 the number  of 
 target nucleons (present in the volume of a cylinder cut along a  
 target diameter  with a 
 radius equal to that of the projectile plus the volumes of the  
 two spherical  segments at the 
 ends of the cylinder \cite {Vid2}) participating in  redistribution 
 of the initial  kinetic 
 energy. We note that similar relationship between the stopping   
 power and  projectile size, 
 for a fixed target, also exists at AGS  energies \cite {Vid2}. 
 This, apparently,  is  not the 
 case at SPS energies since the stopping is found to be  
 independent of the  projectile size 
 \cite {Abe}, suggesting a different stopping mechanism. 

       In conclusion, the rapidity distributions are examined in 
  C+C, d+C, C+Ta  and  d+Ta 
 collisions at 4.2A GeV/c, in  terms of their Gaussian  components. These 
 last are  interpreted as  isotropic 
 thermal sources (fireballs) in relative motion. Additionally, 
 the   corresponding amount of 
 stopping is analysed as a function of system size and  collision centrality. 
 We found that 
 three Gaussians are obtained in  all peripheral  collisions and 
 also in C+C central 
 collisions, two in case of d+C and  C+Ta central  collisions and, 
 finally, one Gaussian in 
 case of d+Ta collisions.  The corresponding  values of the average 
 rapidity loss are 
 calculated and they support the  qualitative relationship between the 
 number of Gaussian 
 components  in the rapidity spectra and the  corresponding stopping power. 
 It is observed 
 that three Gaussians describe  the nuclear  transparency, 
 while single Gaussian 
 corresponds to the full stopping.  Also, we find   in central 
 collisions,  that  the average rapidity loss 
 increases with   the 
 ratio of     the number of target and the number of 
 projectile participants.\\

   ACKNOWLEDGMENTS
 
   The authors would like to thank the members of the collaboration that participated in 
 data processing, and I. Menda\v {s} for valuable comments.

\clearpage

\begin{table}
\caption {Average rapidity loss  $\langle \delta y \rangle$ in
 peripheral and central  (3-5\% cut) collisions.}
\label{tab1}

\begin{tabular}{c c c c c c c c c c c c c} 
\multicolumn{1}{c}{Colliding system}&
\multicolumn{1}{c}{Peripheral collisions}&
\multicolumn{1}{c}{Central collisions}\\ \hline
&\multicolumn{1}{c}{$\langle \delta y \rangle$}&
 \multicolumn{1}{c}{$\langle \delta y \rangle$}\\ \hline

$^{12}$C+$^{12}$C     & $0.53\pm 0.01$ & $0.62\pm 0.01$\\
$^{2}$d +$^{12}$C     & $0.57\pm 0.01$ & $0.86\pm 0.02$\\
$^{12}$C+$^{181}$Ta   & $0.56\pm 0.01$ & $1.14\pm 0.01$\\
$^{2}$d +$^{181}$Ta   & $0.63\pm 0.01$ & $1.42\pm 0.02$\\

\end{tabular}
\end{table}

 \begin{figure}
 \mbox{\epsfig{file=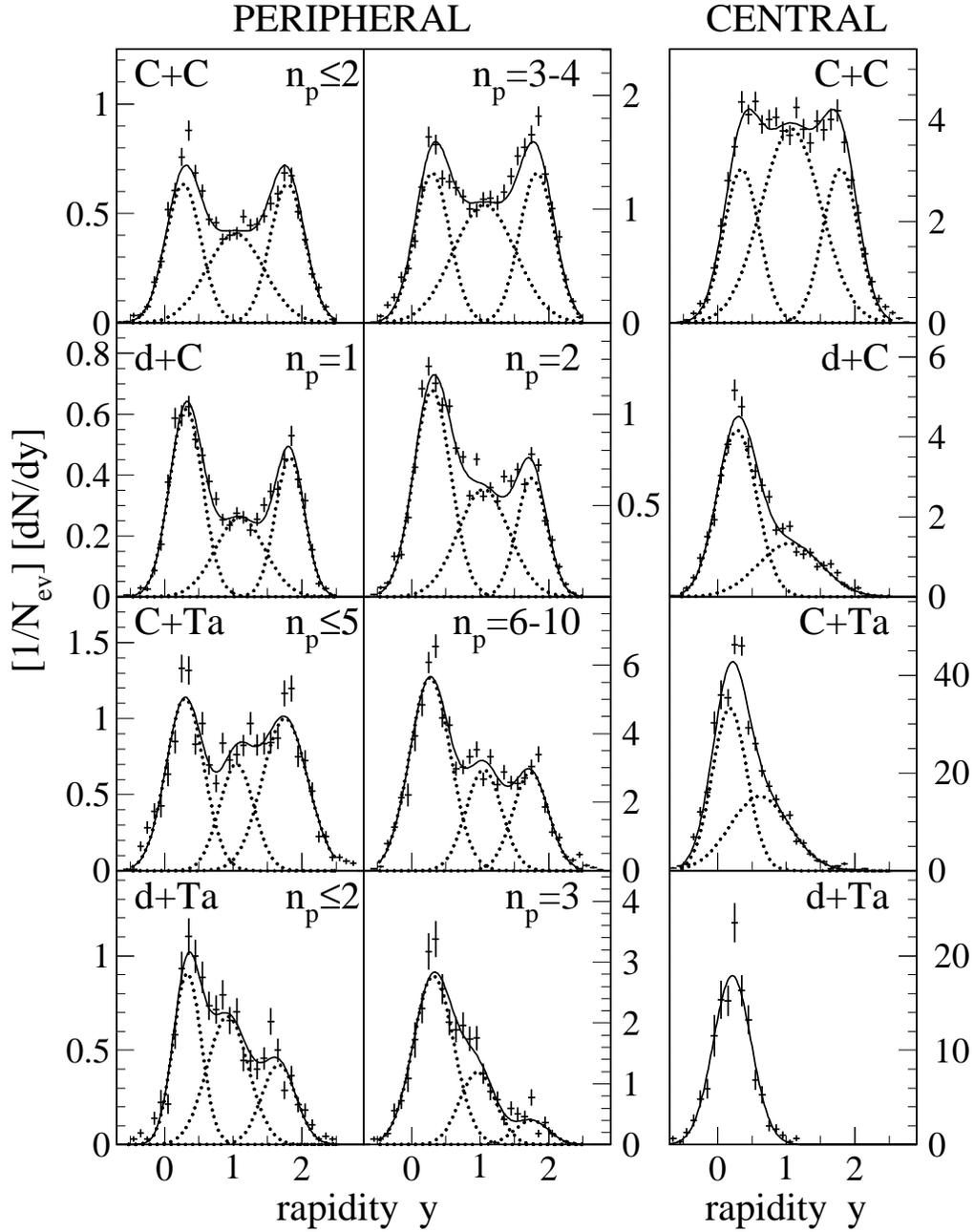,width=14cm}} 
 \caption{
 Rapidity distributions of participant protons in peripheral and central 
 collisions. The solid lines represent best fit to a sum of 
 Gaussians (doted lines).} 
 
 \label{fig1}
 \end{figure}

\end{document}